# Holographic surface measurement system for the Fred Young Submillimeter Telescope

Xiaodong Ren[a*], Pablo Astudillo[b], Urs U. Graf[a], Richard E. Hills[c], Sebastian Jorquera[b], Bojan Nikolic[c], Stephen C. Parshley[d], Nicolás Reyes[e,f], Lars Weikert[a]

[a]I. Physikalisches Institut, Universität zu Köln, Köln, 50937, Germany;
[b]Departamento de Astronomía, Universidad de Chile, Santiago, 7591245, Chile;
[c]Cavendish Laboratory, University of Cambridge, Cambridge, UK;
[d]Department of Astronomy, Cornell University, Ithaca, NY 14853, USA;
[e]Max Planck Institute for Radio Astronomy, Bonn, Germany;
[f]Departamento de Ingeniería Eléctrica, Universidad de Chile, Santiago, 8370451 Chile.

## ABSTRACT

We describe a system being developed for measuring the shapes of the mirrors of the Fred Young Submillimeter Telescope (FYST), now under construction for the CCAT Observatory. "Holographic" antenna-measuring techniques are an efficient and accurate way of measuring the surfaces of large millimeter-wave telescopes and they have the advantage of measuring the wave-front errors of the whole system under operational conditions, e.g. at night on an exposed site. Applying this to FYST, however, presents significant challenges because of the high accuracy needed, the fact that the telescope consists of two large off-axis mirrors, and a requirement that measurements can be made without personnel present. We use a high-frequency (~300GHz) source which is relatively close to the telescope aperture (<1/100th of the Fresnel distance) to minimize atmospheric effects. The main receiver is in the receiver cabin and can be moved under remote control to different positions, so that the wave-front errors in different parts of the focal plane can be measured. A second receiver placed on the yoke provides a phase reference. The signals are combined in a digital cross-correlation spectrometer. Scanning the telescope provides a map of the complex beam pattern. The surface errors are found by inference, i.e. we make models of the reflectors with errors and calculate the patterns expected, and then iterate to find the best match to the data. To do this we have developed a fast and accurate method for calculating the patterns using the Kirchhoff-Fresnel formulation. This paper presents details of the design and outlines the results from simulations of the measurement and inference process. These indicate that a measurement accuracy of ~3μm rms is achievable.

## 1. INTRODUCTION

The Fred Young Submillimeter Telescope (FYST), formerly known as CCAT-prime, is currently in fabrication before shipment to the Atacama region of northern Chile for installation on Cerro Chajnantor at an altitude of 5600m. The telescope has a clear aperture of 6m diameter and the optics are based on the two-mirror crossed-Dragone[1,2] design, which provides zero blockage, low cross-polarization and a very large field of view (FoV). Additional coma-correction terms[3] are applied to its two mirrors to further improve its diffraction-limited FoV to 26deg$^2$ at 2mm wavelength and 4.4deg$^2$ at 350μm. The dry high-altitude site offers superb observing conditions with routine access to the submillimeter atmospheric windows between 850μm and 350μm and opens up the possibility of observing in the 200-μm (1.5THz) window. To preserve the performance of the telescope at these high frequencies, the surface accuracy is required to be better than 10.7μm rms, with a goal of < 7.1μm under good conditions, e.g. stable temperature and moderate winds.

Although the mirrors will be set to the correct shape with high precision in the factory and then shipped to the site fully assembled, we regard it as essential to have an accurate method of measuring them under operational conditions. We plan to use microwave holography[4,5] for this. The holography technique, which involves measuring the complex field pattern of the antenna and deriving the surface errors from that, has a well-proven record for the measurement of large high-frequency radio telescopes, but applying it to FYST presents several significant challenges:

*ren@ph1.uni-koeln.de Telephone: +49 (0)221 470 3549

1. The requirement on surface accuracy is for the whole telescope system, including manufacturing errors and items like deformations due to the changing orientation of the telescope and environmental temperature. We expect the results from the holography measurements to be used for the final adjustment of the mirror surfaces, so any errors in the measurement will also contribute to the final error. We cannot therefore allow them to take up more than a small fraction of the overall error budget. We have adopted a goal of 3μm rms for the measurement accuracy, where this should include both random errors, e.g. those due to noise or atmospheric fluctuations, and systematic errors due, for example, to inaccurate modelling of electromagnetic effects. This is, to the best of our knowledge, a significantly higher accuracy than has been reported for such measurements thus far.
2. The crossed-Dragone optical layout consists of two large off-axis mirrors. The conventional approach to holography[6] employs a simple inverse Fourier transformation of the (complex) pattern of the antenna. This provides a map of the deformations of the wavefront at the aperture, but it cannot discriminate between the errors contributed by different components in the optical system. A new technique is therefore required to break the degeneracy in the two-mirror system.
3. Since the measurements will often be made at night, when it may not be possible to have personnel present at the telescope because of the high altitude, it must be possible to operate all aspects of the system remotely.

We address the first issue by using a high frequency (295.74GHz) and putting the source relatively close to the telescope (~300m away). Using a higher frequency means that a given fractional error in measuring the pattern converts into a smaller error in the surface and also reduces effects like diffraction from the edges of the panels out of which the mirrors are made. In addition, it enables us to use narrow beams for the source and the receiver which reduces the problems of scattering and reflection of the signals by the ground or other objects. The short path to the source minimizes atmospheric effects and gives us a high signal-to-noise ratio (SNR), which means that receiver noise produces small errors in the surface, even with measurement times as short as a few milliseconds per point.

To solve the degeneracy problem and recover the surface errors on both mirrors individually, we have developed a new method we call 'multi-map' holography, described in the next section. This requires us to make beam maps with the receiver located at several different positions in the focal plane. Because of the third challenge in the list above, we have designed an automated system, which will be mounted behind the focal plane in the receiver cabin, to transfer the receiver between the designated positions.

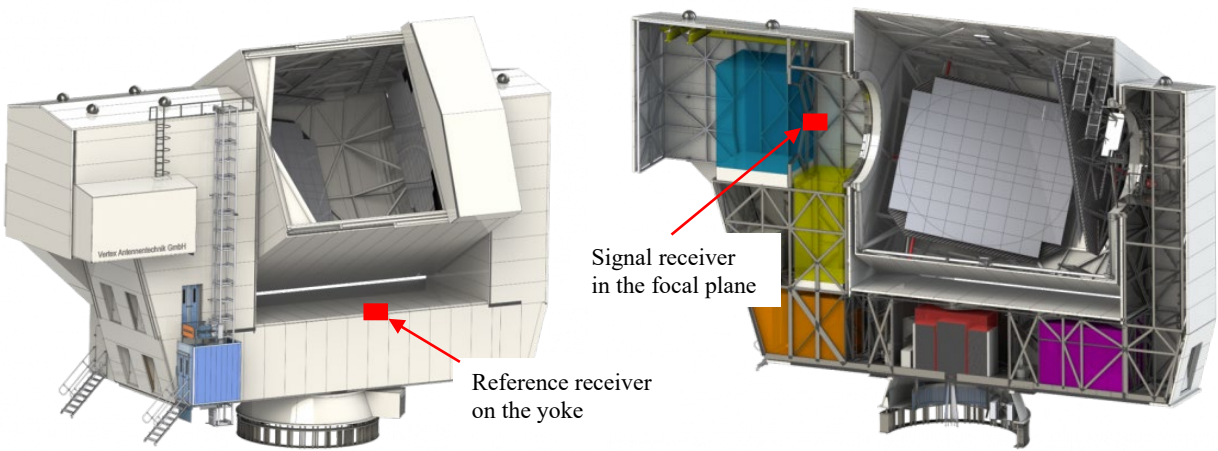

Figure 1. Renderings of FYST showing the locations of the signal and reference receivers.

Since both amplitude and phase of the radiation field are required, a second receiver will be placed on the yoke of the telescope (see figure 1) and used to provide a reference. This observes the source continuously and shares a common microwave reference for its local oscillator (LO) with that of the main receiver. Before correlating the intermediate-frequency (IF) signals from the two receivers, we pass them through a high-resolution digital filter (actually an FFT spectrometer). This greatly improves the SNR and reduces the effects of noise in the source and the LO. All these sub-systems are designed to satisfy the stringent requirements that have been set for the project and to withstand the hostile environment at the site.

In this paper, we describe the FYST holography system, including the innovative analysis approach and the details of the hardware, and we present results from numerical simulations which indicate that the required accuracy of ~3μm should be achievable.

## 2. HOLOGRAPHIC METROLOGY FOR A TWO-MIRROR SYSTEM

The holographic method is based on the fact that the far-field pattern of a telescope is the Fourier transform of the field in the aperture. If, therefore, one measures the complex far-field and makes an inverse Fourier transform, one gets the complex aperture field. For a perfect antenna with the receiver at the nominal focus, the phase in the aperture plane would be constant, so a plot of the measured phase provides a map of any errors in the telescope optics, projected onto the aperture plane. In the standard approach[6], it is usually assumed that the errors in the secondary mirror can be neglected and any errors in the phase for the receiver feed can be corrected (either from theory or laboratory measurements). Corrections can also be applied[7] for the effects of having the source much closer than the far-field distance and the receiver at a displaced position with respect to the focal plane (e.g. defocused). The crossed-Dragone (CD) design is however composed of two large reflectors (primary M1 and secondary M2) of comparable size, as shown in figure 2. Clearly a single beam-pattern measurement cannot discriminate between the phase errors produced by the two mirrors. It could be argued that this does not matter since one could adjust one mirror, say M1, until the phase in the aperture was flat, which would provide a perfect beam even if there were errors in M2 which were being compensated by the deviations which had been put into M1. This is not correct for two reasons: 1) the mirrors are made of relatively large panels (roughly 0.7m square) and the projections of these onto the aperture plane do not match, so an exact compensation is not possible; and 2) the whole point of the CD design is to provide a wide field of view and the compensation would only work at one position in the field – as one moves to other positions the projection of the mirror surfaces onto the aperture plane changes. The converse of this is that, if we measure the beam-pattern with the receiver at several different positions, the degeneracy will be broken and we should be able to obtain separate maps of the errors in the two mirrors. It should, however, be noted that this separation will only work really well for errors with relatively small spatial scales: errors on larger scales will still be nearly degenerate. This can be seen by considering the limiting case of a linear phase error – a tilt of one mirror would be compensated almost exactly by the reverse tilt of the other. It is nevertheless true that, if we explore a sufficiently large area of the focal plane and adjust the mirrors to get good beam patterns at all positions, then we will have achieved our goal.

In principle moving the holography receiver to three different well-separated points in the focal plane would be sufficient to break the degeneracy, but using more points will improve the results. Our simulations indicate that using five points – the center of the field and the corners of a square, as illustrated in figure 2, provides a good compromise between accuracy and the time needed to make a complete set of measurements.

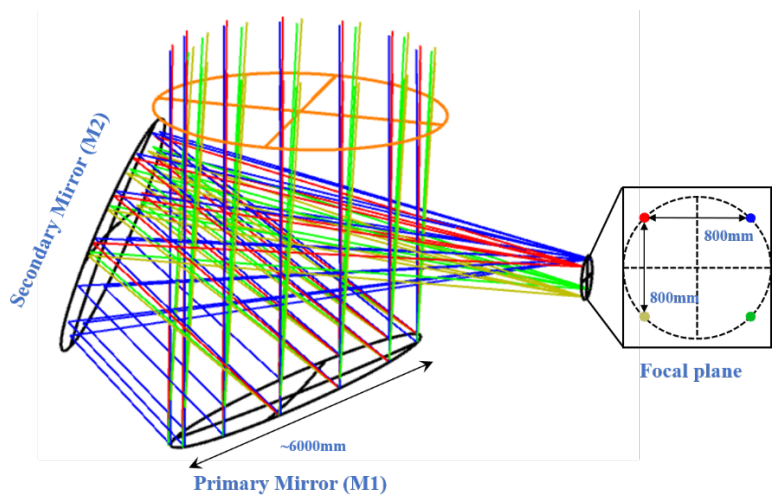

Figure 2. The modified 'crossed-Dragone' optics of FYST is composed of two ~6m reflectors. The ray trace shows the degeneracy between errors on M1 and M2. The ray tracing also indicates that moving the receiver to different well-separated points in the focal plane can change the reflection points on M2 breaking the degeneracy between the mirrors.

There is no obvious direct method, analogous to an inverse Fourier transform, for converting the five measured beam patterns into two maps of the errors in the mirrors. Instead we treat this as a numerical inference problem. Specifically, we find the set of movements of the support points of the individual panels in both mirrors that can best account for the measured data. This requires a fast and accurate algorithm for calculating the "forward" problem – the beam patterns to be expected for a given set of movements. There are numerous advantages with this approach: 1) we can use an accurate formulation of the electromagnetic aspects that includes a proper treatment of the rather complex geometry of the telescope and of the diffraction that takes place between the components and in the path from the source; 2) we can include various systematic effects such as the uncertainty in the position of the receiver, the illumination pattern produced by the feed and telescope pointing errors; and 3) we can control the magnitudes of the terms (such as the tilts of individual mirrors discussed above) where degeneracies remain. The remainder of this section describes the method we have devised to do all this.

## 2.1 Parameterization of the surfaces and other errors

The mirrors consist of panels, 77 on M1 and 69 on M2, with panel sizes of 670 x 750mm and 700 x 710mm respectively. Each panel is supported by five adjustable mounts, one located in the center and the four others at the corners of a rectangle, 440 x 500mm in size. Since we wish to be able to set the mirrors, rather than simply measure them, what we actually need are the values of the adjustments that should be made to obtain a perfect surface. We therefore choose these adjustment values to be the parameters that we vary to find the best fit to the beam-pattern data. With five adjusters we can control five surface error terms for each panel - piston, x or y tilt, curvature, and twist. This means that we should use a second-order polynomial describing these forms of error to interpolate between the support points onto a grid suitable for performing the electromagnetic calculations. As an alternative, which may be useful for the diagnosis of some issues such as thermal deformations, we can ignore the panels and their mounting points and parameterize the whole of each mirror in terms of a set of orthogonal functions such as Zernike polynomials[8]. The maximum order of the polynomials is determined by the spatial resolution required.

There are a number of effects other than the surface error in the mirrors which will cause a mismatch between the model and the data. These include an offset in the pointing of the telescope, bulk misalignments of the mirrors, an error in the position of the receiver and an error in the assumed distance to the source. So long as these errors are small, they all create either a slope or a curvature in the phase at the telescope aperture plane. We cannot, therefore, find the magnitudes of the effects separately from the data. Fortunately, we do not need to know these for the purpose of setting the mirrors, but we must allow for them in the model. We do this by including in the model a term, given by equation 1, for the phase at the aperture plane. Here $x$ and $y$ are the coordinates of a point in the aperture plane and $p$, $q$ and $r$ are free parameters. Note that these parameters will have slightly different values for each of the beam patterns.

$$\Delta\phi_{aperture}(x, y) = p \cdot x + q \cdot y + r \cdot (x^2 + y^2) \tag{1}$$

A term of this form could, however, also be produced by certain combinations of movements of the adjusters in either of the mirrors. This would give rise to degeneracies in the solution when seeking the best fit. We avoid these by using a regularization term, as described below.

We also need to account for the possibility that the amplitude pattern in the aperture is not an exact match to the one that we have assumed, which is based on electromagnetic models of the feed horn and the source. For small deviations, this can be treated in a similar manner to the phase, with a second-order polynomial which multiplies the amplitudes. Here an error in the alignment of the feed would result in a linear term, while an incorrect estimate of the beam-width would produce a quadratic one. In this case there is no issue of degeneracy with the panel movements.

## 2.2 Modelling the antenna radiation fields

In the actual experiment, the signal propagates from the distant source, through the two telescope mirrors and into the receiver. It is more convenient, however, to model the equivalent time-reversed process, starting from the receiver and through the optics, to find the field on a spherical surface at the distance of the source. (The pattern is mapped by scanning the whole telescope in azimuth and elevation, so this ~300m diameter sphere should be centered on the crossing point of the axes.) The standard approach for such a calculation is physical optics (PO) integration[9]. In our case this involves finding the currents excited in M2 by the field radiated from the receiver feed, then calculating the fields at the surface of M1 produced by those currents in M2, and hence finding the currents in M1, and finally calculating the radiating fields that the currents on M1 produce at the required spherical surface. The PO method is accurate and reasonably fast for systems whose optical components are no more than a few hundred wavelengths in aperture. In our problem, however, the mirrors are more than 6000 wavelengths in diameter, which makes the calculation very time consuming. To provide a "gold standard" model we have nevertheless carried out this simulation using the TICRA GRASP software, which is a standard commercial software package used to analyze reflector antennas. It took around 10 days to calculate a single pattern using a machine with 120 CPU cores.

By contrast, we need a procedure for calculating the patterns that is fast enough to run within an iterative loop but which maintains sufficient accuracy for our quite-demanding application. We have found a way of doing this by incorporating the following changes of approach:

1. Instead of the full PO expressions, we use Fresnel-Kirchhoff diffraction theory[8] where the field is described as scalar quantity, so polarization information is lost, and terms which fall off faster than one over the distance are dropped.
2. We model the propagation from M2 to M1 in two steps[10], going first from M2 to the intermediate focal plane and from there to M1, see figure 3. This greatly reduces the density of points needed to represent the fields at the two mirrors.
3. We exploit the fact that the errors in the mirror are small compared to the overall dimensions, so we only need to do the full calculation once to set up the problem and can then leave out many of the time-consuming steps on subsequent iterations.
4. We use automatic differentiation to provide efficient evaluation of the gradient of the fit with respect the parameters.

We tested the effects of the approximations by comparing the results to those obtained with GRASP. In particular, we experimented to find what density of points on the mirrors are required and how large an area in the auxiliary plane needs to be included. With these values chosen to give adequate accuracy for our purposes (errors ~70dB below the peak) the full calculation takes about 2 minutes per beam pattern on a standard laptop. More importantly we can perform ~1 iteration per second when fitting ~700 parameters to five 2,500-point patterns. Using a GPU provides a further increase in speed of nearly a factor of 10. The details of this work, including an evaluation of the errors introduced by the approximations, will be given in a forthcoming paper.

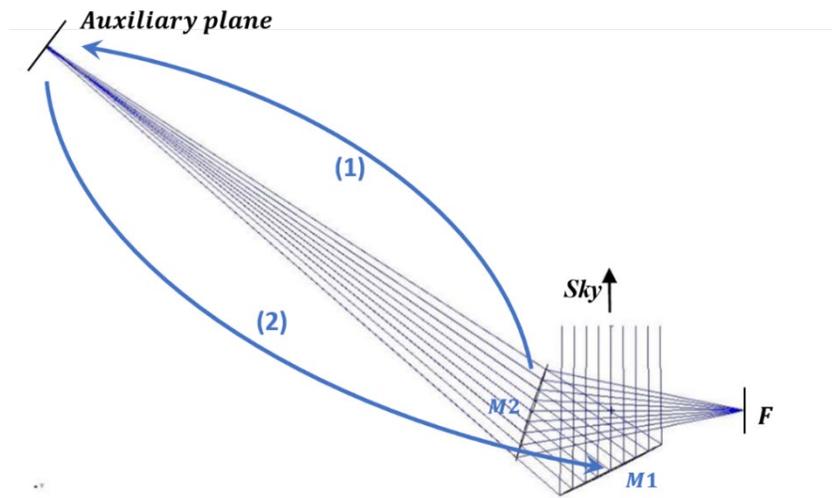

Figure 3. Illustrating the calculation process for the two-step 'Fresnel-Kirchhoff' technique. The field at M2 produced by the feed F is calculated straight-forwardly. To obtain the field at M1 we proceed in two steps, first finding the field at the imaginary intermediate focus – the auxiliary plane – and then back-propagating the field from there to M1. Because we only need to know the final pattern on the "sky" over a limited range of angles, it is sufficient to calculate the field in the auxiliary plane over a relatively small area. This means that far fewer points are needed to describe the fields at M2 than would be the case if we went directly from M2 to M1. The calculation of the beam pattern at the source from the field at M1 is again straight-forward.

## 2.3 Fitting procedure

The data $D_i$ consist of the complex field measured at a set of points $i$ and normalized to have unit total power and zero phase at the origin. We calculate the model fields $Y_i$ as described above, assuming initially that the surfaces are perfect, and apply the same normalization. The mismatch between the data and the model is characterized by the vector differences $E_i = Y_i - D_i$. We wish to find the set of parameters – the movements $S_n$ and $S_m$ of the adjusters on M1 and M2 and the other terms described in section 2.1 above – which minimizes the mismatch, i.e. the sum of $|E_i|^2$ over all the data points. To avoid the degeneracies between large scale terms and to ensure that any linear or defocus terms are not included in the adjuster movements, we include in the quantity to be minimized the magnitudes of the adjuster movements multiplied by the regularization factor $\lambda$, the value of which is not very critical: it needs to be large enough to suppress the degeneracies but not so large that it prevents the method from finding real deformations in the surfaces. The quantity to be minimized is therefore that given by equation 2.

$$R = \sum_i (|Y_i - D_i|)^2 + \lambda \cdot \left( \sum_n S_n^2 + \sum_m S_m^2 \right) \tag{2}$$

The fitting process is illustrated in figure 4.

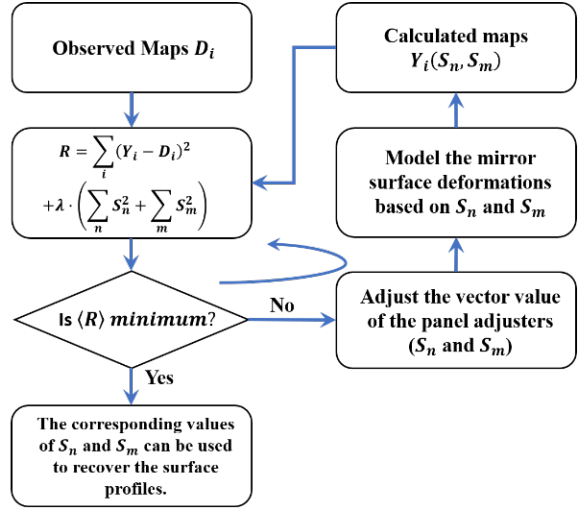

Figure 4. Flowchart of the inference process used to find the adjuster movements required to set the surfaces.

## 3. HARDWARE DESCRIPTION

We now describe the design of the hardware needed to make the measurements. There are four main items: 1) the source which will be mounted on a small tower on the shoulder of the mountain, about 20m higher than the telescope and 300m away; 2) the signal receiver, which is supported by a frame close to the focal plane; 3) the reference receiver mounted on the yoke, immediately below the telescope aperture, where it has a direct view of the source; and 4) a correlation receiver housed in the electronics space inside the yoke. The specifications of the source and receivers are summarized in table 1.

Table 1. Specifications of the holography hardware

|  | Source |  | Receiver | |
|---|---|---|---|---|
|  |  |  | Reference | Signal |
| Frequency | 295.74 GHz | LO Frequency (x2) | 295.68 GHz | 295.68 GHz |
| Beam FWHM | 1.8° | Beam FWHM | 1.1° | 16° |
| Antenna gain | 34 dB | Antenna gain | 40 dB | 84 dB |
| Frequency stability | 1 x 10$^{-9}$ | Noise temperature | < 5800 K | < 5800 K |
| Ouput power | >10 uW | IF Frequency | 60 MHz | 60 MHz |
| Phase noise @ 1kHz | -80 dBc/Hz | Bandwidth | 70 MHz | 70 MHz |

The source and the receivers are all enclosed in temperature-controlled boxes to provide stability and protection from the cold environment. The LO signals for the two receivers are derived from a common microwave reference at 12.32GHz which is also located in the electronics space. Since coherence during the measurement is paramount, identical high-quality and phase-stable cables will be used to carry this reference signal to the two receivers.

### 3.1 Feed horns

The same design of profiled diagonal horn[11] was used for all three RF modules. The horn of the signal receiver illuminates the telescope directly, whereas the source and the reference receivers contain additional optics to provide narrow well-defined beams, which serves both to increase the power level of the signal and to reduce unwanted reflections.

This horn design was chosen because it is relatively simple to machine but has good properties for our application. The half-power beamwidth is 16°, which was chosen to provide an edge taper of 7dB on the telescope. This relatively modest taper enables us to measure the outer parts of the mirrors accurately. The sidelobes are more than 20 dB below the co-polar maximum and the peak cross-polar components are below –25 dB. We have manufactured the horns and verified their compliance with the predictions of electromagnetic modelling by measuring the pattern at 296 GHz in the laboratory, using the planar near-field technique. The reconstructed far-field pattern is shown in figure 5.

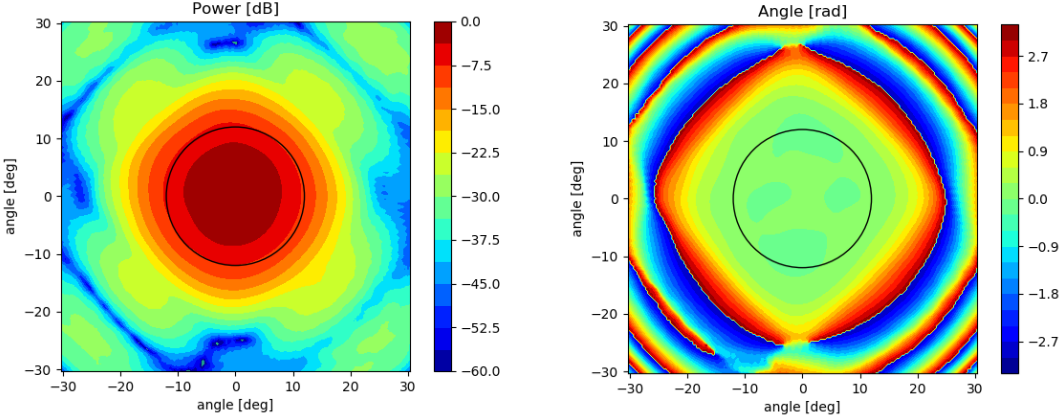

Figure 5. Radiation pattern of the conical-spline horn. The black circle represents the edge of the 6m telescope aperture.

### 3.2 Source

The 295.74GHz signal is produced by multiplying a 16.43GHz dielectric resonance oscillator in two stages – the first active and the second passive. The DRO is in turn locked to a GPS-disciplined 10MHz oscillator. This combination is designed to provide a stable signal with low phase noise. The components making up the source are shown in figure 6. All these components were obtained from commercial sources.

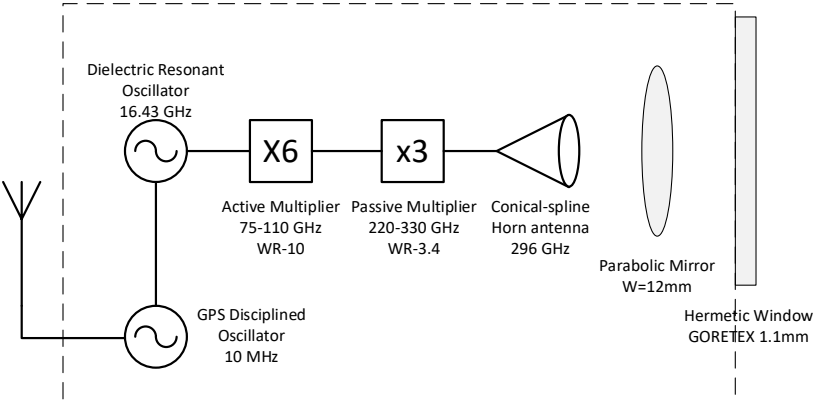

Figure 6. Schematic of the source.

The feed horn illuminates a small off-axis parabolic mirror producing a beamwidth of 1.8° FWHM, which means that it is ~10m wide when it reaches the telescope. This beam size is chosen as a compromise between making it small to minimize the chance of reflections from the ground between the source and the telescope and making it large to ensure that the wavefront across that telescope aperture is spherical. After reflection off a flat mirror, which is used to align the beam, the signal passes through a hermetic window, which is tilted to avoid reflections. Figure 7 shows the layout of the source.

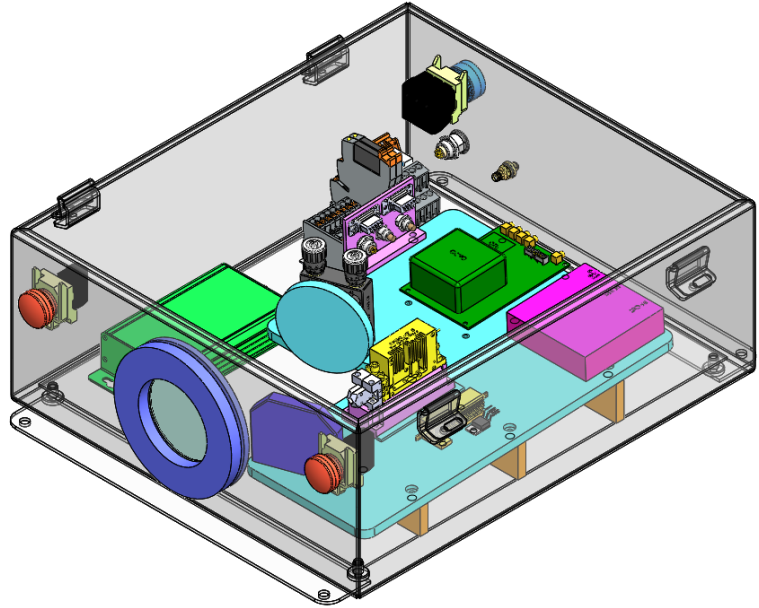

Figure 7. View of the source.

The source will be in an exposed location and will be subjected to low environmental temperatures. The steel enclosure is therefore tightly sealed and lined with thermal insulation. The sensitive components are mounted on a plate which is thermally isolated from the enclosure and thermally stabilized by a heater. The source operates from 24VDC, which is provided by car batteries. These have sufficient capacity for overnight operation, i.e. up to ~12 hours, which will enable us to monitor the effects of temperature variations on the surface errors in the telescope. Since the source will be located beside the road going up to the summit, so that access is relatively easy, no provision for remote control of the source has been made at this point.

### 3.3 Common receiver modules

The two receivers contain almost-identical RF modules. The main components are shown in figure 8.

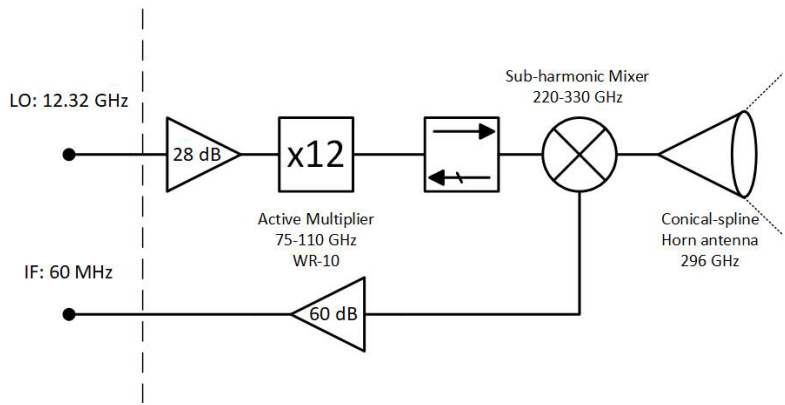

Figure 8. Schematic of the receiver modules.

Each module contains a horn, a sub-harmonic mixer, an active multiplier chain for the LO, and amplifiers for the LO and IF signals. Apart from the horn, these are again commercially available devices. The LO frequency is 147.84GHz, which means that the IF frequency is 60MHz. The signal and reference receiver modules differ only in the amount of IF gain, which is determined by the peak signal level expected. The use of an ambient-temperature subharmonic mixer means that the noise temperature is relatively high, ~5,000K but, because we have a relatively powerful source that is quite close and there is a substantial amount of gain in the optics, this is sufficient to give a signal to noise ratio of more than 70dB.

For this application, simplicity and stability are more important than sensitivity. To assist in this, the components are again mounted on an isolated, temperature-controlled plate and enclosed. See figure 9.

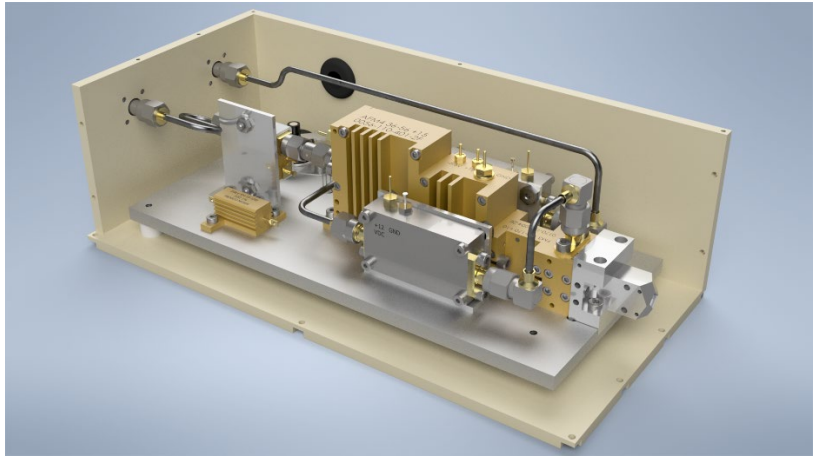

Figure 9. View of receiver module with part of the cover removed.

### 3.4 Signal Receiver

The receiver module to sample the signal in the focal plane is supported by a stiff frame which will be erected in the instrument space when measurements are to be undertaken, see figure 10. The receiver module sits on one of a set of mounting points on this frame, chosen to give the required coverage of the focal plane.

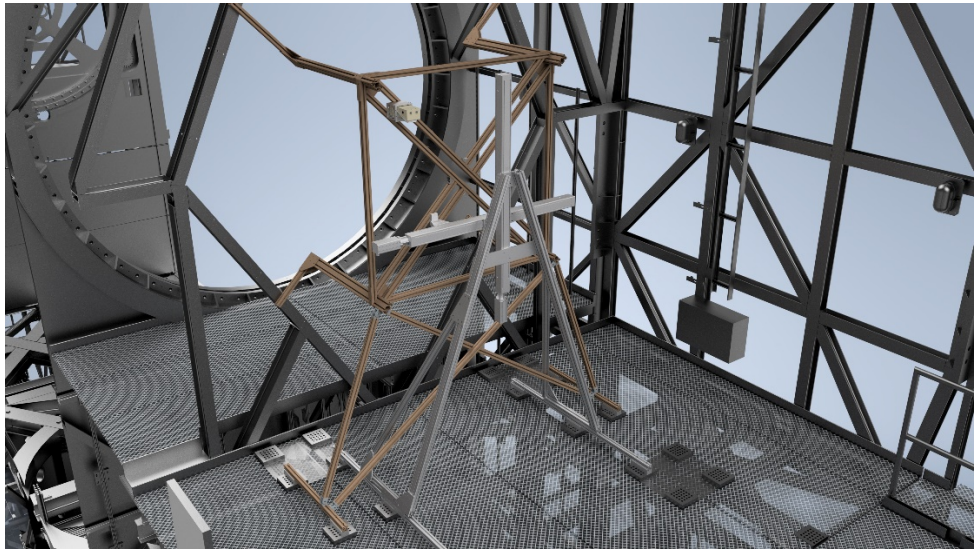

Figure 10. The mount for the receiver in the telescope focal plane, showing the support frame (brown) and the X-Y stage (light grey).

There is a motorized X-Y drive behind the frame that can pick up the receiver module and move it from one mounting point to another under remote control. This should take less than a minute.

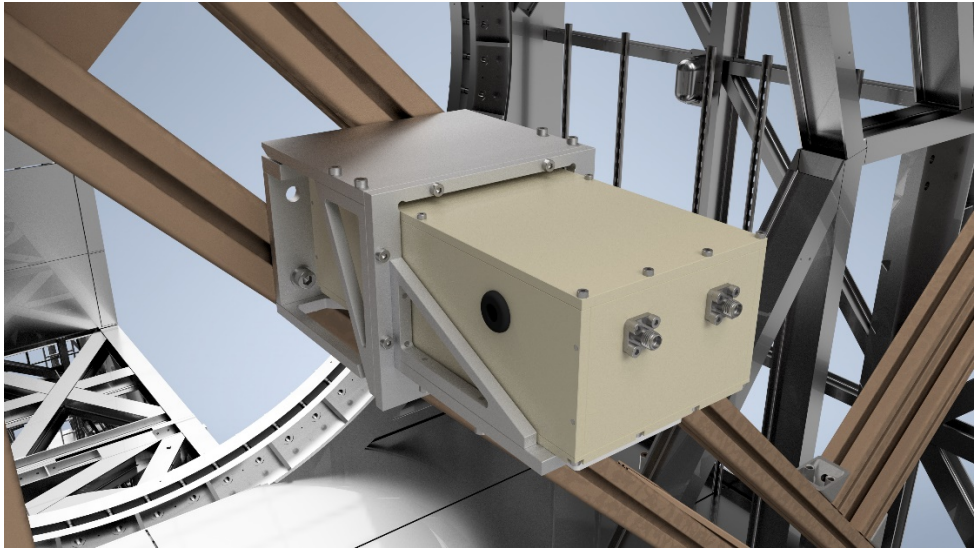

Figure 11. Close-up view of the signal receiver module at one of the mounting positions.

The support of the receiver module is kinematic to ensure that it can be moved and replaced with repeatability. The locations of the mounting points can be changed (manually) if we find that a different spacing is necessary. In addition, the mounting frame can be moved nearer or closer to the focal plane to change the amount of defocus.

### 3.5 Reference receiver

The reference receiver, which will be mounted directly on the structure of the telescope yoke to ensure a stable attachment, contains the second receiver module and optical components, see figure 12.

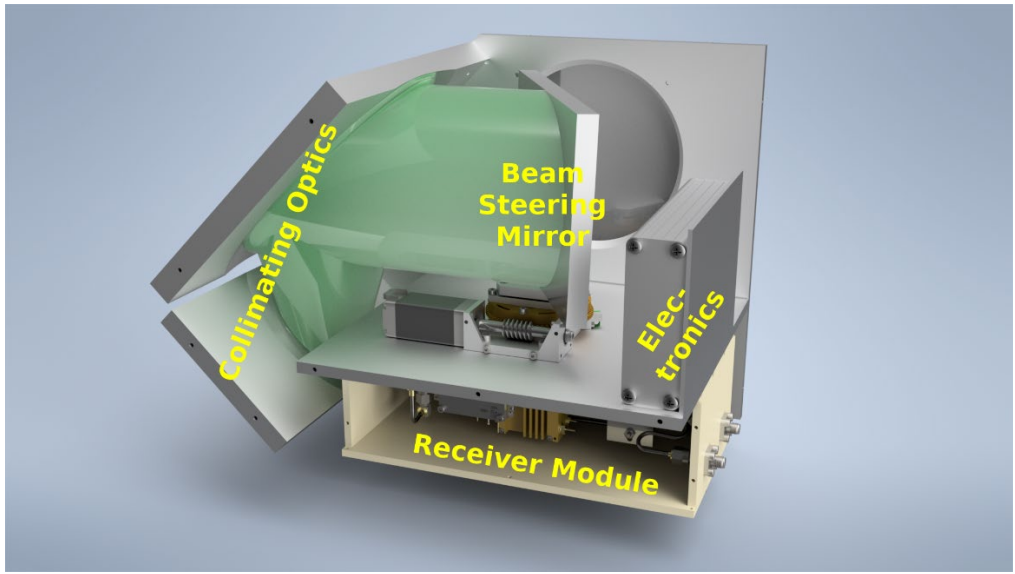

Figure 12. Cut-away view of the reference receiver.

Two off-axis mirrors produce a collimated beam with a FWHM of 1.1°. This is again a compromise between making the beam narrow, to reduce reflections and increase the signal level, and making it wide so that the change in phase and amplitude during the scanning movement, which will typically be ~+/-0.4°, is small. A third flat mirror is mounted on a motorized rotation stage which directs the beam at the source. The rotation is controlled remotely over an Ethernet connection. This is needed because the telescope azimuth at the center of the scan will change when we move the signal receiver to a different position in the focal plane (by about 1.6° for a 400mm offset). We can also make a small scan with this stage, while keeping the telescope stationary and pointed at the source, to calibrate the phase and amplitude of the reference system as a function of azimuth. The mirror can also turn the beam in the opposite direction so that we can make measurements with the telescope flipped "over the back", which will be important for testing whether the deflections in the telescope mirrors due to the force of gravity are consistent with the designers' predictions.

### 3.6 Back-end: a vector-voltmeter with high spectral resolution

We need to measure the amplitude and phase of the signal that has come via the telescope mirrors relative to that from the reference receiver. Since we are using a coherent source it is advantageous for us to make this measurement using a narrow bandwidth. We do this using digital rather than analogue techniques which provides more flexibility and, most importantly, guarantees that the signal and reference are treated in an identical way.

Our approach is to use a dual-input high-resolution spectrometer, essentially acting as a narrow-band filter on the signal and reference, followed by digital multiplication and integration stages. This has been implemented in the Reconfigurable Open Architecture Computing Hardware 2 (ROACH2) platform, composed of two 8-bit high speed analogue to digital converters (ADCs), a Virtex6 FPGA and a PowerPC 440 microprocessor. Figure 13 shows the configuration of this subsystem.

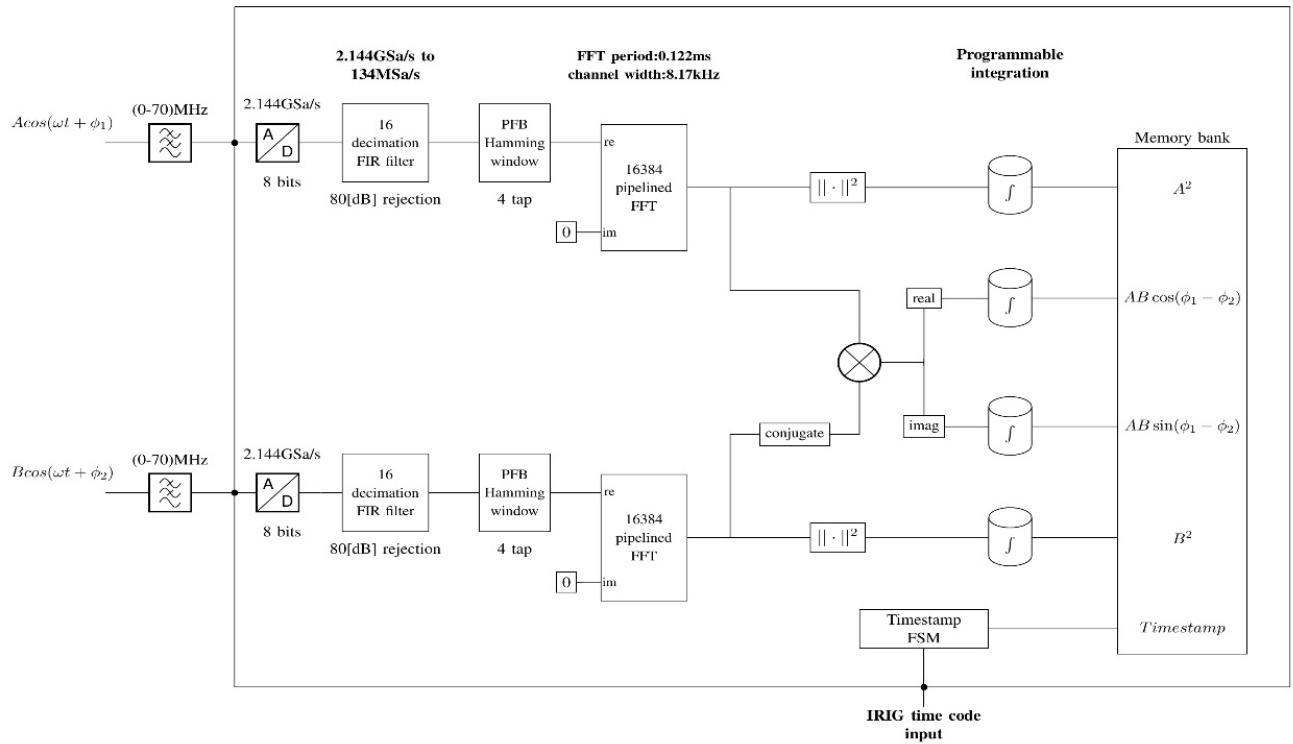

Figure 13: Diagram of vector voltage meter subsystem. The architecture consists in a FX correlator with an oversampling and decimation stage. Concurrently a timestamp subsystem based on the IRIG protocol is running.

We run the ADCs at a high rate, ~2GS/s, and then reduce the sampling frequency by a factor of 16 using a polyphase decimation FIR filter with 80dB of rejection, which gives us a usable bandwidth of 67MHz and reduces the quantization noise by ~12dB[12]. The resampled data are then passed through a pipelined FFT using a polyphase window[13], which provides high isolation between spectral channels and a relatively flat frequency response within a channel. We fine-tune the ADC sampling frequency to place the signal in the middle of a channel. We then select the real and imaginary outputs from this channel and form the cross-product and the magnitudes. These are produced at a rate of ~8kHz and integrated for a selectable time, typically between 1 and 20ms. The data are stored in the internal memory of the PowerPC for post-processing offline.

Concurrently, we run a finite-state machine to provide timestamps so we can align the data samples with the corresponding telescope pointing position. The timestamp subsystem is calibrated using an Inter-Range Instrumentation Group (IRIG) time code signal fed by the observatory's master clock and uses each pulse of the IRIG packet to update its internal value, ensuring that the system is locked to the master clock with a time resolution of 957ns.

We have run numerous tests on this system, of which one is illustrated in figure 14. Here we compare the results from our system with those from a commercial vector network analyzer (Keysight E8364C VNA). We use two signal generators locked to the same frequency, keeping one at a fixed power level (to represent the reference in the holography measurement) while making a sweep in the power level of the other (to represent the signal). Note that the relative phase of the two outputs will alter as the power level is changed because of the switching of the internal attenuators: this is seen in the results – figure 14c. The signals from the two oscillators are split so that we can make measurements simultaneously with the ROACH2 system and the VNA. With the ROACH2 system we collect 128 measurements and calculate averages and standard deviations producing the blue curves in the plots. The red curves are the VNA values.

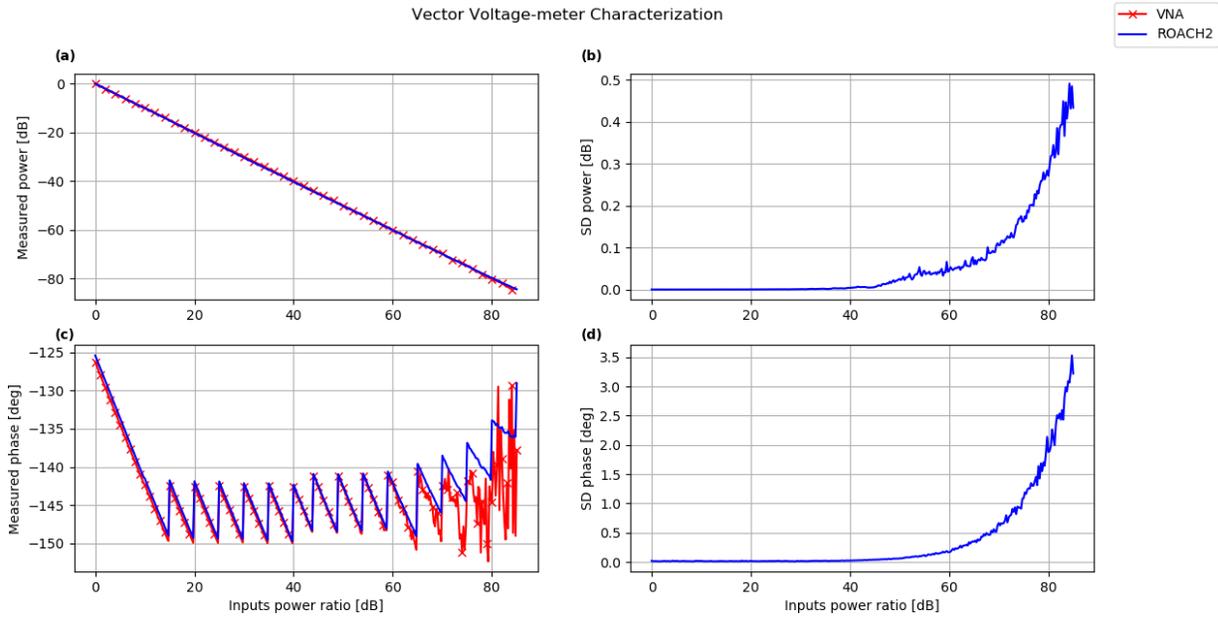

Figure 14: Characterization of the vector voltage meter system. ROACH2 measurements in blue and VNA measurements in red, for input difference in the range (0-85) dB. (a) Average power difference; (b) ROACH2 standard deviation for the power differences; (c) Average phase difference; (d) ROACH2 standard deviation for the phase differences.

We find that the agreement is better than 0.2 dB in amplitude and 1 degree in phase when the ratio of the powers is in the range 0 to 65dB. At lower power levels differences start to appear, but the VNA values become noisy. These results do however confirm the accuracy of our system over a larger dynamic range than we need. The measured standard deviations (figure 14b and 14d) represent an upper limit on internal random errors due to things like round-off in the ROACH2 system. These are well below the levels that would affect our measurements.

## 4. TELESCOPE SCANNING PATTERN, DATA TAKING AND ANALYSIS

To obtain adequate spatial resolution (~0.1m) in mapping the errors in the mirror surfaces, we need to measure the beam pattern over a region about 0.7° in extent. We do this by scanning the telescope in both coordinates while taking data continuously – "on-the-fly" measurements. Note that, because we have adopted the inference approach in our data analysis, we do not need the data points to be on a regular grid and we do not have to have uniform coverage of the whole area. Instead the important criteria are that we can make the maps reasonably quickly, that the pointing positions are known accurately and that we have a way of calibrating the amplitude and phase at frequent intervals. The latter can be accomplished by making the pattern pass repeatedly through the same point in the map, so that any changes in the amplitude and phase can be tracked and taken out of the data.

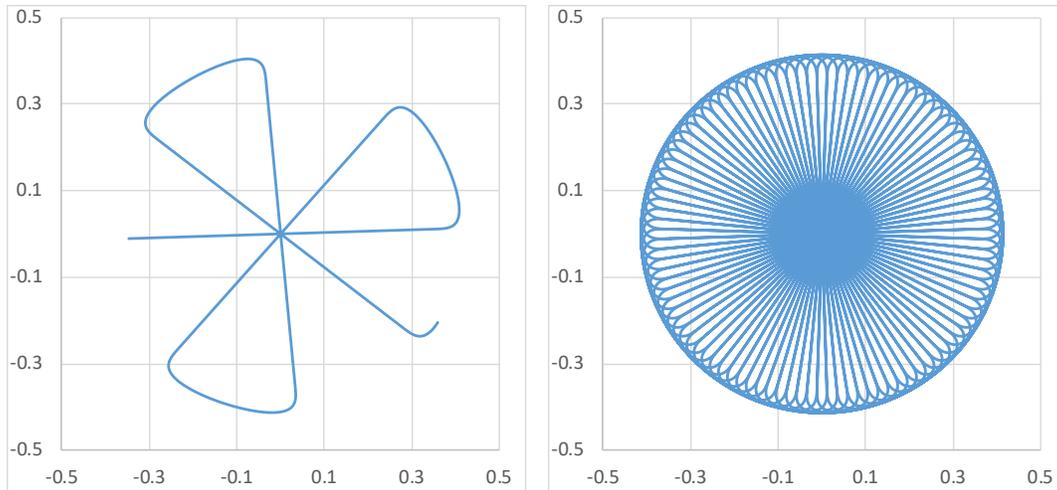

Figure 15. A possible scan pattern showing, left, the first 4 scans and, right, the full pattern with 50 scans. The line is the path of the telescope pointing position and the scales are in degrees.

Figure 15 shows an example of the sort of pattern that we plan to use. This consists of a series of straight lines, all passing through the center point, connected by turn-around loops. Data is taken in the straight sections, where the motion is at constant velocity, so the pointing accuracy should be good. The turns are executed as quickly as is possible without exciting vibrations in the telescope structure and no data is taken during those periods. A pattern with 50 scans samples the beam sufficiently well and this would take 500 seconds to execute if we spend 7 seconds on the straight sections and 3 seconds in the turns. These numbers are conservative in that they require only a small fraction of the full velocity and acceleration capability of the telescope drives. An additional feature of the pattern illustrated here is that successive scans pass through the center point in nearly orthogonal directions. This means that it should be possible to measure, and correct for, any drifts in the telescope pointing that occur on timescales longer than a few scans.

We now outline the steps involved in taking and processing the data:
1. In the case above the scanning rate is ~0.1°/s. The size of smallest angular scales in the pattern is set by the ratio of the measurement wavelength to the telescope aperture and is close to 0.01°. This means we only need about 10 data points per second to sample the pattern fully. To avoid smearing the data along the scan, however, we oversample – e.g. we set the integration time in the back-end to 0.01s giving ~100 points per second. These data are read out, together with their time-stamps, and put in a file.
2. We also read the encoders on the telescope, in this case at a 200Hz rate, again with timestamps which are generated from the same clock as is used for the data. The encoder values are interpolated to find the actual pointing positions at the times when the data samples were taken and these positions are combined with the data.
3. Since the quantity that we need in order to have a result that is independent of amplitude fluctuations in the source is the ratio of the signal to the reference, we divide the values of the voltage product delivered by the back-end by the reference power, i.e. we form $AxB/|B|^2$.
4. The values are corrected for the amplitude and phase response of the reference optics as a function of telescope azimuth. (These are measured in a separate experiment where we scan the mirror built into the reference receiver while keeping the telescope fixed and pointing at the source.)
5. The data is smoothed using a sinc function. The width of this is chosen to make sure that all the real structure in the pattern is preserved but the noise on the samples is reduced as far as possible. It is then resampled at coarser intervals (roughly Nyquist sampling) so that we do not have to process an unnecessarily large number of samples in the later analysis.
6. The data from each scan is interpolated to find the amplitude and phase at the center of the map. Ideally these values would all be the same, but in practice there will be drifts from scan to scan due to changes in the instrument and, probably more significantly, in the atmospheric path between the source and the telescope. (It is of course only the differences between the atmosphere along the path from the source to the telescope aperture and that along the path to the reference receiver that matter, but we cannot expect these to be negligible at the few-micron level that we are concerned about here.) We therefore fit suitable smooth functions of time to the amplitude and phase measured at the center points of the scans and adopt these as representing the drifts that occurred during the course of making the map. We then use these fitted functions to correct the data.
7. We then find the set of surface errors that is most consistent with these measurements using the inference technique described in section 2. Points to note are:
    a. The model has to include the pattern of the feed horn in the signal receiver. While we can fit for the amplitude of this pattern, any error in the phase will be transferred to the mirrors. We do not think that we can measure the horn pattern well enough for our purposes, so we will rely on theoretical modelling of the horn for this. The measurements shown above (figure 5) do confirm that the horn basically works as expected and we are confident that the phase must be very smooth on small and medium angular scales because the physical dimensions of the horn aperture are small in terms of the wavelength. We can check for phase errors on large scales by, for example, making measurements with the horn rotated to different angles about the line of sight. (We will need to use a wire grid to suppress the cross-polar response in this case.)
    b. The processing described above assumes that the pointing of the telescope is stable throughout the course of the map. We can look at the residuals – the differences between the data and the best-fitting models for the beam patterns – to see if there is evidence for pointing errors. If there is, then we can construct a model for these, including possible dependence on time and things like the direction and velocity of motion of the antenna – and fit for these. Corrections to the pointing could then be made and a second iteration of the solution for the surface errors carried out. If really necessary, this "outer loop" could be repeated.
    c. Since the parameters of the fit are the errors at the points where the panels are supported, these values can be used directly to make the adjustments required to bring the surfaces to the desired shape. We would then repeat the measurements. The adjustment is unlikely to be perfect in a single step, especially if the initial errors are large: amongst other things some of the approximations in our model require that the errors be small compared to the observing wavelength. We nevertheless expect the adjustment process to converge after a small number of iterations.
    d. In other cases, where we are simply monitoring the deformations, e.g. as a function of temperature or simply over time, it may be more convenient to fit for the surface shapes as a sum of polynomials. This will give us numerical values for errors of various forms, such as astigmatism and can be used to make images of the surface deformations. Note that, because we are using a source at a fixed elevation, we cannot measure the changing effect of gravity on the structure as the telescope is pointed to different elevations. It is however possible to flip the telescope over and turn the azimuth through 180 degrees so that it again points at the source. This essentially reverses the gravity vector with respect to the mirrors, so we should be able to check whether or not the response to this component of gravity is as expected.

## 5. SIMULATIONS

We have carried out extensive simulations of the measurement and fitting process in order to test the accuracy and sensitivity of the method and to find the optimum set-up. As an example, we need to decide where to locate the receiver with respect to the focal plane of the telescope. Because the source is only 300 meters away from the aperture, the "best" focus, i.e. that giving the peak on-axis gain, is ~705mm behind the astronomical focal plane. Our method of calculating the pattern does not however rely on having the receiver at this "best" focus and it is easy to see that it is in fact better to have it somewhat closer to the focal plane. Figure 16 shows cuts through the beam pattern for the in-focus case and for the case with the receiver moved forwards by 105mm. The effect is to spread out the beam so that the peak gain is reduced by ~20dB. This is advantageous since it means that the dynamic range required in the measurement is reduced by 20dB. We could in principle defocus further but, if we spread the beam too far, we will start to lose some of the detailed information about the surface errors because parts of the pattern will be pushed outside the area being measured. It can be seen that with 105mm of defocus the power is still largely contained within a region ~0.2° in diameter which is small compared to the ~0.7° diameter of the map. Detailed simulations confirm that this amount of defocus gives good results, so we plan to use that initially, but we have designed the mount for the receiver to allow for changes in this offset over the range 0 to 500mm with respect to the "best" focus position in case we need to try different amounts of defocus.

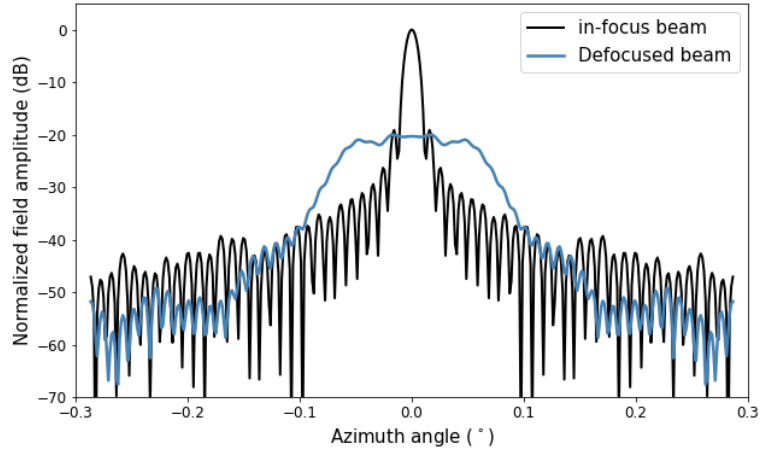

Figure 16. Cuts through the in-focus beam (black) and the defocused case (blue) with the horn 105mm closer to the focal plane than the nominal best-focus position.

The patterns in figure 16 include the full geometry of the telescope but assume perfect mirror surfaces. To test our data analysis process, we introduce a set of random movements at the panel adjusters with a magnitude to 30µm rms. Figure 17 shows the resulting surface errors.

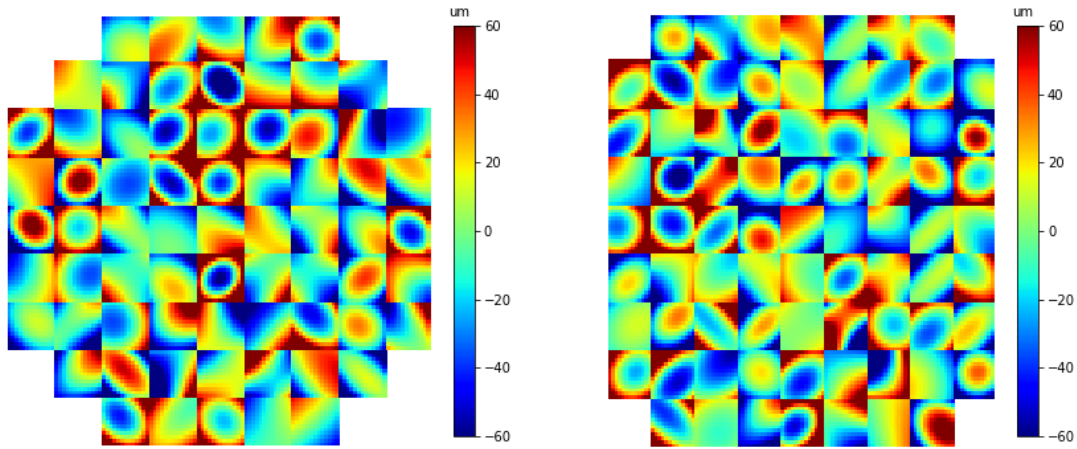

Figure 17. Surface deformation maps with 30µm rms random movements at the adjusters: secondary mirror (M2) on the left and primary mirror (M1) on the right.

For the simulations we use four positions of the receiver at the corners of a square which is 800mm on a side. The resulting beam patterns, with and without the surface errors, are shown in figure 18. If we take these four maps and feed them into our iterative fitting process (taking perfect surfaces for the starting point), we recover the same set of adjuster movements to high accuracy (<<1 µm). This is not surprising since we are using the same method to calculate the beam patterns and to do the forward calculation in the fitting. (As explained in section 2, we have separately checked our calculation method by comparison with the industry-standard GRASP software.)

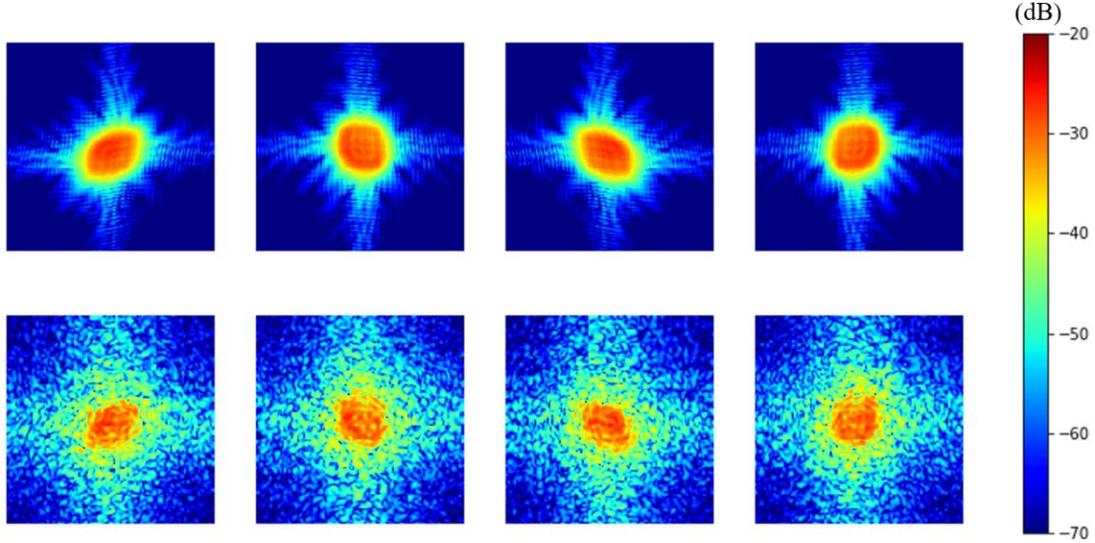

Figure 18 Amplitude of the beam patterns (relative to the in focus peak) with the receiver located at four different points in the focal plane, from left to right, receiver located at: [400,400], [400,-400], [-400,400] and [-400,-400]. Top row – ideal mirror surfaces; bottom row – with the surface errors shown in figure 17. The squares cover 0.57 by 0.57 degrees.

### 5.1 The effects of the noise in the data

To test the effects of noise in the data we replace the calculated (complex) data values $D_i$ with $D'_i$, given by equation 3.

$$D'_i = D_i \cdot (1 + \Delta G_i)e^{j\Delta\phi_i} + N_i \tag{3}$$

Here $N_i$ is a set of random Gaussian variables with zero mean and rms $\sigma$ on each component, representing additive noise, and $\Delta G_i$ and $\Delta\phi_i$ are multiplicative errors which could arise from gain and phase fluctuations in the equipment or the atmosphere. We define the signal-to-noise ratio (SNR) as the ratio of the peak amplitude, for the case of map made with perfect mirrors and with the receiver at the best-focus position, to the noise, $\sigma$. The multiplicative terms are random but, to make them realistic they are not white noise but instead filtered to produce a smoother variation with time. This enables us to test the process of correcting for the gain and phase variations described above (step 6 in section 4). The details of that work is beyond the scope of this paper: here we outline the most important examples.

Figure 19 shows the deviation of the derived surface from that assumed (i.e. the surface shown in figure 17) for a case with additive white noise and a SNR of 65dB.

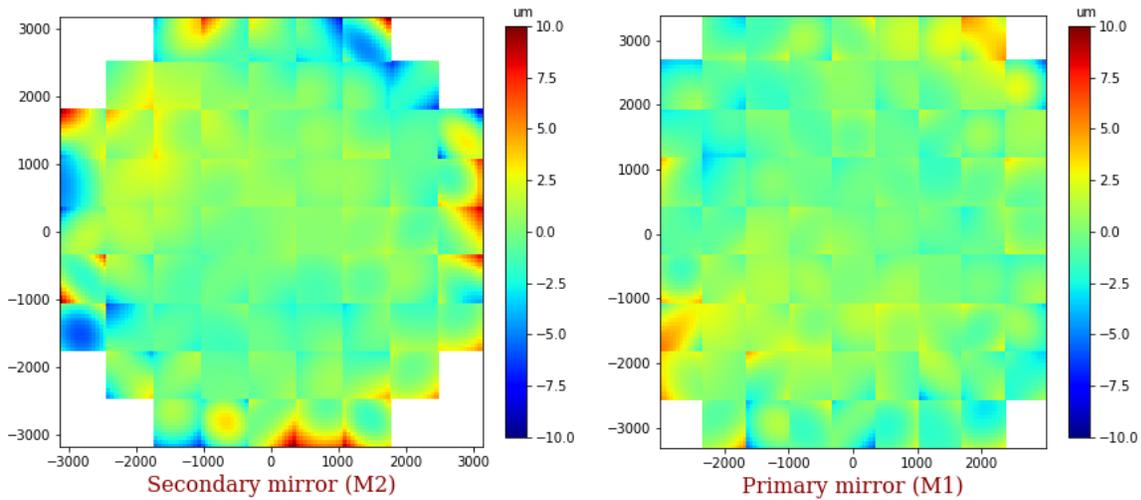

Figure 19. Deviations of the best-fitting surfaces from the input surfaces where noise with amplitude –65dB with respect to the peak power was added to the data. The root-mean-square of the surface error is 1.2μm on M1 and 1.7μm on M2.

In this case the overall root-mean-square of the adjuster differences, $\epsilon$, is around 1.5µm. It can be seen, however, that most of the errors occur near the edges of the mirrors, particularly on M2. In the case of M1 the reason for this is the lower illumination at the outer edge by the receiver horn. On M2, only the inner region contributes to all four of the beam patterns – the regions near the corners are in the signal path for only one of the maps. This means that the rms over the whole surface is a pessimistic estimate of the actual operational performance of the telescope.

Figure 20 shows how the error depends on the signal to noise ratio and on the spacing of the receiver locations in the focal plane. The SNR is expressed in terms of the voltage ratio, so that the range covered is from 55dB (voltage ratio 560) to 70dB (3200). As might be expected, using a small spacing (+/-200mm in each direction) produces rather large errors, while +/-400mm is good and +/-600mm is only a little better. The mount for the receiver allows us to use up to +/-500mm.

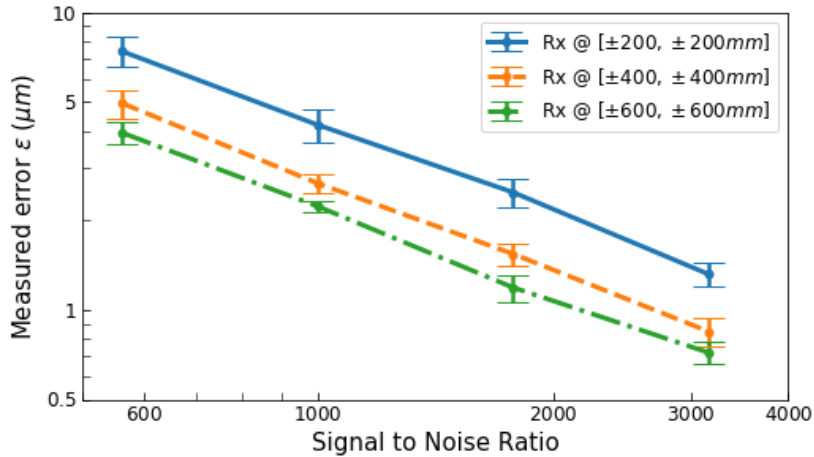

Figure 20. The root-mean-square error $\epsilon$ in deriving the mirror surfaces as a function of signal to noise ratio. The three traces are the results when using different sets of receiver positions as indicated in the legend.

The plot indicates that we need an SNR of greater than 70dB to be sure that the measured errors $\epsilon$ are below one micron. Recall, however, that the SNR is here defined as the ratio of the peak in-focus power to the noise, whereas the case being modelled is for an out-of-focus receiver position where the peak power is reduced by 20dB. Furthermore, the resulting wavefront errors, taking account of the real detector illumination pattern, are substantially lower than the calculated values of $\epsilon$. The sensitivity of our system and the power levels in the signals mean that signal to noise ratio in the data, even with the defocus, will be well in excess of 50dB. This means that additive noise should not have a significant effect on our results.

We also expect the effects of gain and phase fluctuations in the instrumentation to be small. For example we find that random gain or phase variations with a flat spectrum, i.e. white noise, at the level of 1 part in $100 - \langle \Delta G_i^2 \rangle^{1/2} = 0.01$ or $\langle \Delta \phi_i^2 \rangle^{1/2} = 0.01$ radians – each produce surface errors $\epsilon$ of about 0.5 microns. The gain and phase fluctuations in the electronics of our system should be well below those levels.

### 5.2 The effects of atmospheric phase errors

As the signal from the source passes through the atmosphere, the effective path will fluctuate due to variations in the refractive index of the air. At millimeter wavelengths, these variations are mainly due to differences in the amount of water vapour in the turbulent cells, particularly near the ground, but temperature differences may also play a role. For example, even on the high dry site planned for the FYST where the atmospheric pressure is ~0.5bar and the temperature ~265K, the total additional path due to refraction over the 300m distance is about 45mm and, if the relative humidity is 20%, the water vapour contributes about 1mm. This means that either a change in the temperature of ~0.6K or a change in the humidity from 19% to 21% would change the path by ~100µm. The path from the source to the reference receiver is, however, close to that from the source to the telescope aperture, so most of the variations will be common-mode and will not affect the measurements. We expect nevertheless that the remaining fluctuations will often be significant.

To model the atmospheric phase fluctuations, we assume that they are random with a power-law spectrum with a slope of –8/3, which is what is expected if they are produced by Kolmogorov turbulence[14]. We can make a suitable time series by generating white noise, transforming to the frequency domain and adjusting the magnitudes to have the desired slope, and then transforming back to the time domain again. Figure 21 shows the power spectrum of one such series.

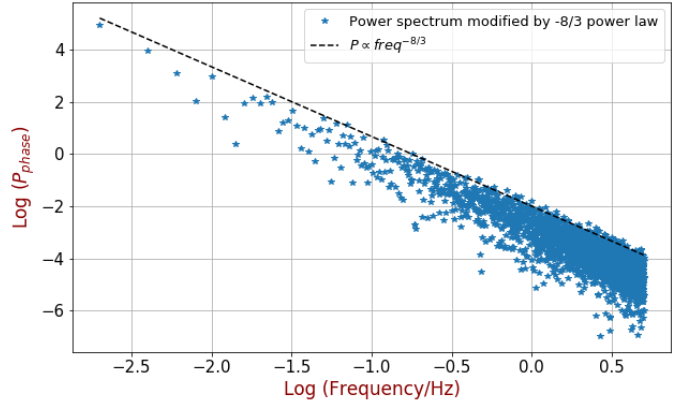

Figure 21. Temporal power spectrum of the phase fluctuations assumed for simulation of atmospheric effects.

Figure 22 shows the time sequence of the phase errors. If we apply these to a model set of data and solve for the mirror surfaces we find that the rms error in the solution $\epsilon$ is 2.5μm.

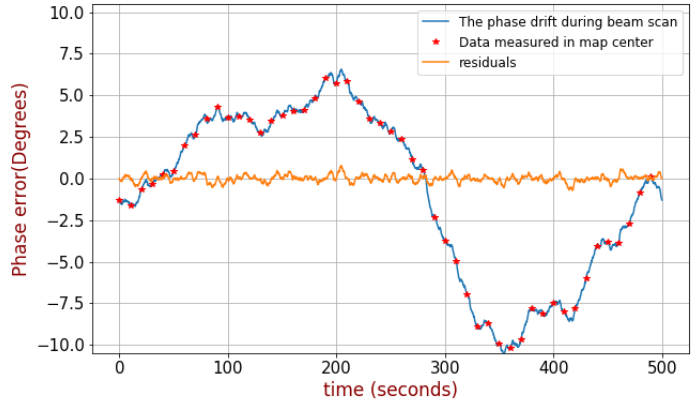

Figure 22. Simulated phase error due to atmospheric variations as a function of time (blue). The data points where the beam passes through the center of the map are in red. The residuals after correction using these data points is in orange.

We can, however, apply the correction procedure described in section 4, where we take out the more slowly varying components by using the data points taken when the beam is passing over the center of the pattern. This leaves the residual errors shown as the orange line in figure 22. With these applied to the model data, we find errors in the solution of only 0.5μm. Although we cannot predict in advance what the magnitude of the atmospheric phase errors will really be at the site, this result demonstrates that the correction technique that we plan to use should make a large reduction in their effects.

## 5.3 The effects of telescope pointing errors

Given that the beam maps contain a lot of structure on angular scales of order the wavelength over the aperture diameter, it is not surprising that the results are quite sensitive to pointing errors. A simulation with "random-walk" pointing errors of 2 arc seconds peak-to-peak in each coordinate gives surface errors $\epsilon$ of 2.5μm rms, which is already higher than we can tolerate. We do not know what the real pointing errors will be, but we expect that they will be low, given that the maps cover a small angular region and only take about ten minutes to complete. As indicated in section 4 (point 7b), we believe that there is sufficient redundant information in the data, particularly near the center of the map where there is a high density of samples, to enable us to find and correct for pointing errors so long as these have a reasonably simple behavior, e.g. a slow drift with time and/or a straightforward dependence on the direction and velocity of motion of the antenna.

## 6. DISCUSSION AND CONCLUSIONS

The design of the holographic measurement system for FYST has been presented. This breaks new ground in both the accuracy that it is expected will be achieved and in the fact that it can measure and separate the errors in the two large mirrors making up the FYST optics. To achieve this, we have: 1) found a way of breaking the degeneracy between the mirrors by making measurements at multiple positions in the focal plane; 2) developed a method for calculating the beam patterns of a complex optical system that is very efficient but still takes proper account of diffraction effects, even when the system is very large in terms of the numbers of wavelengths across the aperture; and 3) designed the necessary hardware and software for carrying out these measurements. The simulations that we have carried out so far indicate that we should be able to reach the required measurement accuracy. We nevertheless anticipate that there will be many practical issues that will need to be resolved when we start taking real data, for example it is quite possible that actual forms of the atmospheric phase errors or of the telescope pointing errors will turn out to be quite different from what we have assumed. We are, however, confident that the basis of the approach set out here is sound.

The various subsystems described here are presently in the final stages of construction. Laboratory testing of the system will start early in 2021 and this will be followed by tests on the telescope when it undergoes trial assembly at the factory.

## 7. ACKNOWLEDGMENTS

It is a pleasure to acknowledge the encouragement and support of the CCAT-prime Observatory. The work at Universität zu Köln is carried out within the Collaborative Research Centre 956, sub-projects D2 and S, funded by the Deutsche Forschungsgemeinschaft (DFG) – project ID 184018867. The team at Universidad de Chile acknowledge the support from ANID through project QUIMAL 180004